\documentstyle[12pt]{article}
\input{epsf.tex}
\begin{document}
\begin{titlepage}
\begin{flushright}
hep-ph/9610271{\hskip.5cm}\\
SISSA/107/96/EP\\
IOA-04-96
\end{flushright}
\begin{centering}
\vspace{.1in}
{\bf THRESHOLD EFFECTS AND RADIATIVE ELECTROWEAK SYMMETRY
 BREAKING IN $SU(5)$ EXTENSIONS OF THE MSSM.} \\
\vspace{1 cm}
{\bf A.Dedes}$^{(1)}$, {\bf A.B.Lahanas}$^{(2)}$,
{\bf J.Rizos}$^{(3)}$
and {\bf K.Tamvakis}$^{(1)}$\\
\vspace{1 cm}
$^{(1)}${\it Division of Theoretical Physics,
Physics Department,}\\
{\it University of Ioannina, GR-45110, Greece}\\ \vspace{0.5cm}
$^{(2)}${\it
University of Athens, Physics Department, Nuclear and
Particle Physics Section,}\\
{\it Ilissia, GR-157 71 Athens, Greece}\\ \vspace{0.5cm}
$^{(3)}${\it
International School for Advanced Studies, SISSA,}\\
{\it Via Beirut 2-4, 34013
Trieste, Italy}

\vskip 1cm
{\bf Abstract}\\
\end{centering}
\vspace{.1in}
We make a complete analysis of radiative symmetry
breaking in the MSSM and its $SU(5)$ extensions
including low- and high-energy threshold effects
in the framework of the two-loop renormalization group.
 In particular, we consider {\it minimal} $SU(5)$, the
{\it missing-doublet} $SU(5)$, a
 {\it Peccei-Quinn} invariant version of
$SU(5)$ as well as a  version
with light adjoint remnants. We derive permitted ranges
for the parameters of these models in relation
to predicted  $\alpha_{s}$ and $M_G$
values within the present experimental accuracy.
The parameter regions allowed under the constraints
of radiative symmetry breaking, perturbativity and proton
stability, include the experimentally designated domain for
$\alpha_s$. In the case of the {\it minimal} $SU(5)$, the
values of $\alpha_s$ obtained are somewhat large in
comparison with the experimental average. The
{\it missing-doublet} $SU(5)$, generally, predicts smaller
values of $\alpha_s$. In both versions of the {\it missing-doublet},
the high energy threshold effects on $\alpha_s$ operate in the
opposite direction than in the case of the minimal model, leading
to small values. In the case of the {\it Peccei-Quinn} version
however the presence of an extra intermediate scale allows to
achieve an excellent agreement with the experimental $\alpha_s$
values. Finally, the last considered version, with light remnants,
exhibits unification of couplings at string scale at the expense
however of rather large $\alpha_s$ values.
\vspace{1.0cm}
\begin{flushleft}
October 1996 \\
\end{flushleft}
\vspace{0.05in}
\hrule
\vspace{0.05in}
E-mails\,:\,adedes@cc.uoi.gr, alahanas@atlas.uoa.gr,
rizos@susy.he.sissa.it, tamvakis@cc.uoi.gr
\end{titlepage}

{\bf 1. Introduction}
\vspace{.5 cm}

Supersymmetric unification, in the framework of  supersymmetric grand unified
theories (SUSY GUT's) $\cite{Nilles}$, or Superstrings $\cite{witten}$, is in
good agreement $\cite{amaldi}$ $\cite{polonsky}$
with the low energy values of the three gauge
couplings, known to the present experimental accuracy, as well as with
available bounds on the stability of the proton. The effective low energy
theory resulting from such a framework is a supersymmetric $SU(3)_c \times
SU(2)_L \times U(1)_Y$, model with softly broken supersymmetry. The simplest
model of that class is the minimal extension of the standard model (MSSM)
$\cite{Nilles}$. A most appealing feature realized in the MSSM is the breaking
of the electroweak symmetry through radiative corrections $\cite{Iban}$. The
Higgs boson mass squared parameters, although positive definite at high
energies, are
radiatively corrected, as can be most easily studied by the use of the
renormalization group, yielding a negative mass squared eigenvalue at low
energies which triggers electroweak symmetry
breaking $\cite{Ross}$ $\cite{castano}$.

In the present article we study the radiative breaking of electroweak symmetry
in the framework of $SU(5)$ SUSY GUTs. In addition to the standard low energy
inputs ($\alpha_{EM}$, $G_F$, $M_Z$, ...) and the soft breaking parameters
($M_o$, $M_{1/2}$, $A_o$), we have the thresholds of the superheavy particles
parametrized in terms of at least two more parameters ($M_{\Sigma}$,
$M_{H_c}, ...$\,)\, . Our output includes the strong coupling $\alpha_s$ and
the unification scale $M_G$, as well as the complete spectrum of new particles.
The predicted strong coupling values can depend strongly on the high energy
thresholds.
Thus, our analysis discriminates between the various GUT models.

Our basic low energy inputs are the boundary values of the gauge couplings in
the $\overline{DR}$ scheme $\cite{anton}$
$\hat{\alpha}_1\equiv\alpha_1(M_Z)|_{\overline{DR}}$
and $\hat{\alpha}_2\equiv\alpha_2(M_Z)|_{\overline{DR}}$. Their values can be
determined in terms of the Fermi constant $G_F=1.16639\times 10^{5}$
GeV$^{-2}$, the Z-boson mass $M_Z=91.1884\pm0.0022$ GeV $\cite{langacker}$,
the electromagnetic coupling $\alpha_{EM}^{-1}=137.036$, the bottom quark mass
$m_b=5$ GeV, the tau mass $m_{\tau}=1.777$ GeV, and the top quark mass.
We can write down the following formulas,
\begin{eqnarray}
\hat{\alpha}_1^{-1}&=&\frac{3}{5} \alpha_{EM}^{-1}\, cos^2\theta\, (1-\Delta
_{\gamma}+\frac{\alpha_{EM}}{2 \pi} \ln(\frac{M_S}{M_Z}))\\[3mm]
\hat{\alpha}_2^{-1}&=& \alpha_{EM}^{-1}\, sin^2\theta\, (1-\Delta
_{\gamma}+\frac{\alpha_{EM}}{2 \pi} \ln(\frac{M_S}{M_Z}))
\end{eqnarray}
where $\Delta_{\gamma}$=0.0682$\pm$0.0007 $\cite{vayonakis}$ includes the
light quark and lepton contributions, and the $\overline{DR}$ mixing
angle is given by
\begin{equation}
sin^2\theta = \frac{1}{2} \{1-[1-\frac{4 \pi \alpha_{EM}}{\sqrt{2} G_F M_Z^2
(1-\Delta r)}]^{\frac{1}{2}}\}
\end{equation}
The quantity $\Delta r$ will be given below. The scale $M_S$ appearing in (1)
and (2) is not a physical scale but a convenient parametrization for the
contribution of all sparticle and heavy particles $(W, t,H^{+})$,
defined as
\begin{equation}
M_S=\frac{M_W^{-7}M_t^{\frac{16}{9}}M_{H^{+}}^{\frac{1}{3}}
M_{\tilde{t}_1}^{\frac{4}{9}}M_{\tilde{t}_2}^{\frac{4}{9}}
M_{\tilde{u}_{1,2}}^{\frac{8}{9}}M_{\tilde{u}^c_{1,2}}^{\frac{8}{9}}
M_{\tilde{b}_1}^{\frac{1}{9}}M_{\tilde{b}_2}^{\frac{1}{9}}
M_{\tilde{d}_{1,2}}^{\frac{2}{9}}M_{\tilde{d}^c_{1,2}}^{\frac{2}{9}}
M_{\tilde{\tau}_1}^{\frac{1}{3}}M_{\tilde{\tau}_2}^{\frac{1}{3}}
M_{\tilde{e}_{1,2}}^{\frac{2}{3}}M_{\tilde{e}^c_{1,2}}^{\frac{2}{3}}
M_{\tilde{\chi}_1}^{\frac{4}{3}}M_{\tilde{\chi}_2}^{\frac{4}{3}}}
{M_Z^{\frac{19}{9}}}
\end{equation}
An additional useful parametrization scale $\tilde{M_S}$, relevant to the
strong coupling constant ${\alpha}_s$, is
\begin{equation}
\tilde{M_S}=\frac{M_t^{\frac{2}{3}}
M_{\tilde{t}_1}^{\frac{1}{6}}M_{\tilde{t}_2}^{\frac{1}{6}}
M_{\tilde{u}_{1,2}}^{\frac{1}{3}}M_{\tilde{u}^c_{1,2}}^{\frac{1}{3}}
M_{\tilde{b}_1}^{\frac{1}{6}}M_{\tilde{b}_2}^{\frac{1}{6}}
M_{\tilde{d}_{1,2}}^{\frac{1}{3}}M_{\tilde{d}^c_{1,2}}^{\frac{1}{3}}
M_{\tilde{g}}^2}
{M_Z^{\frac{11}{3}}}
\end{equation}
The quantity $\Delta r$ appearing in Eq. (3) can be written as
$\cite{bagger}$,
\begin{equation}
\Delta r = \Delta_{\gamma} - \frac{\alpha_{EM}}{2\pi} \log\frac{M_S}{M_Z} -
\frac{\Pi_{ZZ} (M_Z^2)}{M_Z^2}+\frac{\Pi_{WW}(0)}
{M_W^2}+\Delta_{SM}+\delta\rho^{QCD}+\delta\rho^{HIGGS}
\end{equation}
$\Pi_{ZZ}$ and $\Pi_{WW}$ stand for the Z and W self-energies calculated using
dimensional reduction. The quantity $\Delta_{SM}$ stands for Standard Model
vertex+box corrections and is given by $\cite{sirlin}$,
\begin{equation}
\Delta_{SM}= \frac{\alpha_{EM}}{4 \pi
sin^2\theta}\{6+\frac{\log{cos^2\theta_W}}{sin^2\theta_W} [\frac{7}{2}-\frac{5
sin^2\theta_W}{2}-sin^2\theta (5-\frac{3
cos^2\theta_W}{2 cos^2\theta})]\}
\end{equation}
In the last formula, by definition $cos^2\theta_W=\frac{M_W^2}{M_Z^2}$. The
pole mass of W-gauge boson in (4) is related to  $M_Z$ by
$M_W^2=M_Z^2 \,\rho\, cos^2\theta$ where the $\rho$ parameter is given by,
\begin{equation}
\rho=1-\frac{\Pi_{WW}(M_W^2)}{M_W^2}+\frac{\Pi_{ZZ}(M_Z^2)}{M_Z^2}+2-loop\,
finite \,corrections.
\end{equation}
We have explicitly written in (6) the 2-loop corrections due to QCD and the
Higgs calculated in reference $\cite{kniehl}$.

The low energy value of the strong coupling constant considered as an output is
given by the formula $\cite{hall}$,
\begin{equation}
\alpha_s^{-1}\equiv\alpha_s^{-1}(M_Z)|_{\overline{MS}}=
\hat{\alpha}_3^{-1}(M_Z)|_{\overline{DR}}+\frac{1}{4\pi}-
\frac{1}{2\pi} \ln{\frac{\tilde{M_S}}{M_Z}}
\end{equation}
The current experimental values of $\alpha_s$ at Z-pole  extracted from
 QCD experiments with various methods are obtained
 in Table I $\cite{langacker}$. R refers to the ratio of cross sections or
partial widths to hadrons versus leptons and the values of strong coupling
are in $\overline{MS}$ renormalization scheme. The average value of
$\alpha_{s}$ given in
Ref.$\cite{langacker}$ is $\alpha_s(M_Z)=0.118\pm 0.003$.

The values (1),(2) will serve as low energy boundary conditions for the
corresponding two-loop renormalization group equations. As a high energy
boundary conditions for the gauge couplings we shall impose unification at a
scale $M_G$,
\begin{equation}
\hat{\alpha}_1(M_G)=\hat{\alpha}_2(M_G)=\hat{\alpha}_3(M_G)\equiv \alpha_G
\end{equation}
Note that when we depart from the minimal supersymmetric extension of the
Standard Model and consider $SU(5)$ extensions of it the effect of the
superheavy particles with masses around the unification scale $M_G$ has to be
taken into account in the evolution of the gauge couplings.

\begin{center}
\begin{tabular}{cc}\hline
\multicolumn{2}{c}{\bf TABLE I} \\ \hline
\multicolumn{2}{c}{$Process \: \;\;\;\;\;\;\;\;\;\;\;
\;\; \: \;\;\;\;\;\;\;\;\;\;\;\;\;\;\;\;\;\;\;\;\;\;\;    \alpha_s(M_Z)$ }
\\
Deep inelastic scattering  &0.112$\pm$ 0.006 \\
R in $\tau$ lepton decay        &0.122 $\pm$ 0.005  \\
R in $\Upsilon$ decay   &0.108 $\pm$ 0.010  \\
Event shapes in e$^+$e$^-$ annihilation &0.122$\pm$ 0.007  \\
$Q\overline{Q}$ lattice         &0.115$\pm$ 0.003 \\
Fragmentation   &0.122$\pm$ 0.012  \\
Jets at HERA    &0.121$\pm$ 0.012  \\
R in Z$^o$ decay (LEP and SLC)  &0.123$\pm$ 0.006  \\
Deep inelastic at HERA  &0.120$\pm$ 0.014  \\
\hline \hline
\end{tabular}
\end{center}
\centerline{\footnotesize {\bf Table I:} Values of $\alpha_s(M_Z)$ extracted
from QCD experiments}
\vspace{1cm}

The soft supersymmetry breaking is represented by four parameters $M_o$,
$M_{1/2}$, $A_o$ and $B_o$ of which we shall consider only the first three as
input parameters and treat $B(M_Z)$, as well as the Higgs mixing parameter
$\mu$, as determined by the 1-loop minimization equations,
\begin{eqnarray}
sin2\beta&=&-\frac{2 B \mu}{\overline{m}_1^2+\overline{m}_2^2}\\[3mm]
\frac{1}{2} (M_Z^2+\Pi_{ZZ}(M_Z^2))&=&\frac{\overline{m}_1^2-
\overline{m}_2^2 tan^2\beta}{tan^2\beta-1}
\end{eqnarray}
where $\overline{m}_i^2\equiv m_{H_i}^2+\mu^2+\frac{\partial(\Delta
V_{1-loop})}{\partial \upsilon_i^2}$. Note that cases in which the above two
equations are not satisfied, and thus radiative symmetry breaking does not
occur, are rejected. The parameter $\beta\equiv
tan^{-1}(\frac{\upsilon_2}{\upsilon_1})$ is defined at $M_Z$. $M_Z$ in (12)
denotes the experimental  Z-boson mass.
Thus supersymmetry breaking is
parametrized with the input parameters $M_o$, $M_{1/2}$, $A_o$, $\beta(M_Z)$
and $sign\mu(M_Z)$. We shall take the simplest of boundary conditions at $M_G$
assuming universality,
\begin{equation}
m_i(M_G)=M_o\, \, ,\,\,M_i(M_G)=M_{1/2}\,\,,\,\,A_i(M_G)=A_o
\end{equation}
We shall employ the full coupled system of 2-loop renormalization group
equations $\cite{martin}$ evolved from $M_G$ down to low energies. Since our
purpose is to study the effect of high energy thresholds in various extensions
of $SU(5)$ it will be sufficient to obtain the parameter $M_S$ and
$\tilde{M_S}$,
appearing in the boundary values of the low energy gauge couplings, by
calculating the sparticle masses in the
step approximation $\cite{dedes}$. However, we shall
introduce finite part contributions whenever they are {\it a priori}
expected to be large $\cite{zhang}$, such as $QCD$ corrections to the top
and gluino masses for instance etc.

Following the Particle Data Group $\cite{langacker}$ our basic experimental
constraints on supersymmetric masses as well as Higgs boson masses
are shown in Table II. These limits, impose stringent bounds on the
extracted values of $\alpha_s(M_Z)$ and on heavy high energy masses as we
 will see later.
\begin{center}
\begin{tabular}{cc}\hline
\multicolumn{2}{c}{\bf TABLE II} \\ \hline
\multicolumn{2}{c}{$Particle \: \;\;\;\;\;\;\;
\;\; \: \;\;\;\;\;\;\;\;\;\;\;\;\;\;\;\;\;\;    Bound (GeV)$ }
\\
Neutralinos  & \\
$m_{\chi_1^o}$ (LSP)    &$>$23  \\
$m_{\chi_2^o}$  &$>$52  \\
$m_{\chi_3^o}$  &$>$84  \\
$m_{\chi_4^o}$  &$>$127 \\
Charginos       &  \\
$m_{\chi_1^c}$  &$>$45.2 \\
$m_{\chi_2^c}$&$>$99  \\
Sneutrino       &  \\
$m_{\tilde{\nu}}$ &$>$41.8 \\
Charged Sleptons &  \\
$m_{\tilde{e},\tilde{\mu},\tilde{\tau}}$ &$>$45 \\
Squarks & \\
$m_{\tilde{q}}$ & $>$224 \\
Gluino & \\
$m_{\tilde{g}}$ & $>$154 \\
Higgs Bosons & \\
$m_h$ & $>$44 \\
$m_A^o$ & $>$24.3 \\
$m_H^{\pm}$ &$>$40 \\
\hline \hline
\end{tabular}
\end{center}
\centerline{\footnotesize {\bf Table II:} Experimental bounds on supersymmetric
particles and Higgs Bosons}
\vspace{1cm}

The purpose of the present paper is to analyze the effect of high energy
thresholds on unification predictions in the various $SU(5)$ extensions of the
MSSM.
Our goal is to
complete previous
existing analyses which either have not
incorporated the constraints imposed by radiative symmetry breaking or
have not considered the full range allowed for GUT parameters.
In our consideration of supersymmetric versions of $SU(5)$ we have to
take into account the constraints imposed by proton decay into
$K^+\overline{\nu}_{\mu}$ through dimension 5 operators.
Assuming that this is the dominant process the proton lifetime is
$\cite{nath}\cite{hisano} \cite{wright}$
\begin{equation}
\tau(p\longrightarrow K^{+} {\overline{\nu}}_{\mu})=6.9\times10^{31}yrs
|\frac{0.003 GeV^3\, 0.67\, \sin(2\beta)\, M_{H_c}\, TeV^{-1}}{\beta_n\, A_S
(1+y^{tK})\, 10^{17} GeV\, [f(\tilde{u},\tilde{d})+f(\tilde{u},\tilde{e})]}|^2
\end{equation}
where $M_{H_c}$ is the effective colour triplet mass in GeV,$\beta_n \sim
(0.003 - 0.03) GeV^3$,  $|1+y^{tK}|\geq 0.4$, and
$A_S=(\frac{\alpha_{1}(M_Z)}{\alpha_{5}(M_G)})^{\frac{7}{99}}
(\frac{\alpha_3(M_Z)}{\alpha_5(M_{GUT})})^{-\frac{4}{9}}$. The function
$f(x,y)$ is defined as
\begin{equation}
f(x,y)=m_{\tilde{w}}\frac{1}{x^2-y^2}
(\frac{x^2}{x^2-
m_{\tilde{w}}^2}\ln(\frac{x^2}{m_{\tilde{w}}^2})-
\frac{y^2}{y^2-
m_{\tilde{w}}^2}\ln(\frac{y^2}{m_{\tilde{w}}^2})\,)
\end{equation}
and masses are in GeV. The current experimental limit is
\begin{equation}
\tau(p\longrightarrow K^{+}{\overline{\nu}}_{\mu})\geq  10^{32} yrs
\end{equation}
In what follows, we choose the most conservative values of $\beta_n$
and $|1+y^{tK}|$ which are $0.003 GeV^3$ and $0.4$ respectively.

An additional constraint that will be imposed is the absence of Landau poles on
the dimensionless couplings of the theory, or equivalently the validity of
perturbation theory (perturbativity)
of those couplings above $M_G$ and up to the Planck scale. Although this is
in general not a severe constraint, it should be taken into account in the
cases of extended versions of $SU(5)$ due to the existence of a large number
of massless particles above $M_G$ (see for instance Fig. 3c).
This constraint is implemented through the
numerical solution of the 1-loop RG equations for the Yukawa couplings of
$SU(5)$ above $M_G$ (see Appendix).
\vspace{.5 cm}

{\bf   2. Minimal SU(5)}
\vspace{.5 cm}

{}The standard superpotential \cite{dimopoulos} in the minimal $SU(5)$ is
\begin{eqnarray}
\cal W &=& \frac {1}{2} M_1 Tr(\Sigma^2)+\frac{1}{3}\lambda_1
Tr(\Sigma^3)+M_2\overline{H} H+\lambda_2 \overline{H}\Sigma H \nonumber
\\[1.5mm]&+&\sqrt{2}
Y_{(d)}^{ij}\Psi_i\phi_j\overline{H}+\frac{1}{4}Y_{(u)}^{ij}\Psi_i\Psi_j
H \;\;\;\;\;\; i,j=1..3
\label{pot}
\end{eqnarray}
$SU(5)$ is spontaneously broken to $SU(3) \times SU(2) \times U(1)$
when the adjoint Higgs $\Sigma$ develops a v.e.v in the direction
$<\Sigma>\equiv V\,Diag(2,2,2,-3,-3)$. The resulting superheavy masses are ,
\begin{equation}
 M_V=5  g_5 V \,\, , \,\, M_{H_c}=5 \lambda_2 V \,\, , \,\, M_{\Sigma}=5
\lambda_1 V
\end{equation}
We have imposed the usual fine-tuning condition $M_2=3\lambda_2 V$ that gives
massless isodoublets of $H$,$\overline{H}$. The mass $M_{\Sigma}$ stands for
the mass of the surviving colour-octet and isotriplet parts of $\Sigma$.
The leading contribution of these masses to the beta function coefficients of
the three gauge couplings $\frac{dg_i}{dt}=\frac{g_i^3 (b_i+\Delta b_i)}{16
\pi^2}, i=1...3$, $t=ln(\frac{Q}{M_G})$, $(b_3,b_2,b_1)=(-3,1,\frac{33}{5})$
is
\begin{eqnarray}
\Delta b_3&=& 3\theta(Q^2-M_{\Sigma}^2)+
\theta(Q^2-M_{H_c}^2)-4\theta(Q^2-M_{V}^2)\} \\[2mm]
\Delta b_2&=& 2\theta(Q^2-M_{\Sigma}^2)-
6\theta(Q^2-M_{V}^2) \\[2mm]
\Delta b_1&=& \frac{2}{5}
\theta(Q^2-M_{H_c}^2)-10\theta(Q^2-M_{V}^2)
\end{eqnarray}
\begin{figure}
\hbox to\hsize{\hss \epsfysize=3in
  \epsfbox[18 18 552 462]{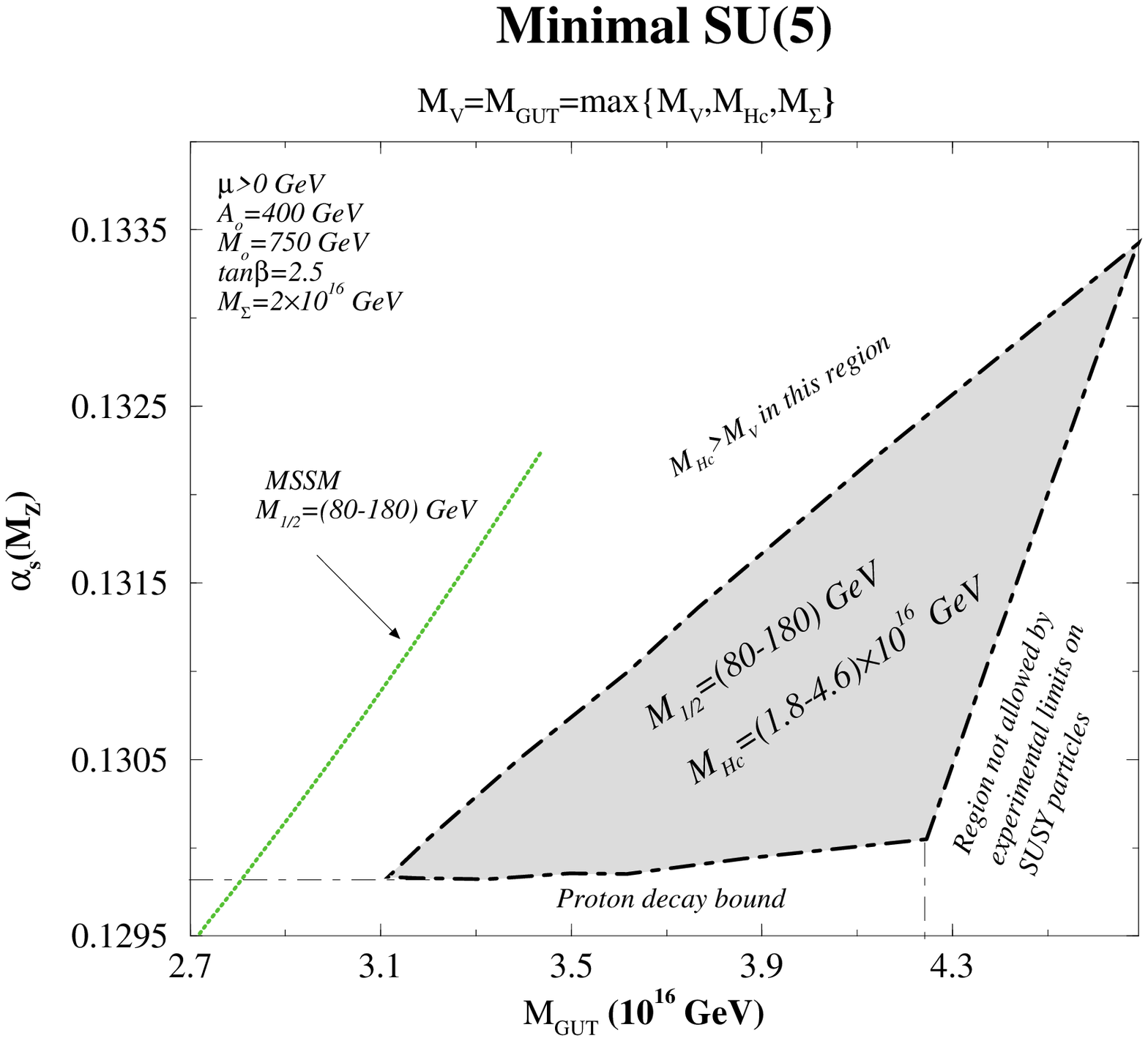}\hss
   \epsfysize=3in\epsfbox[18 18 552 462]{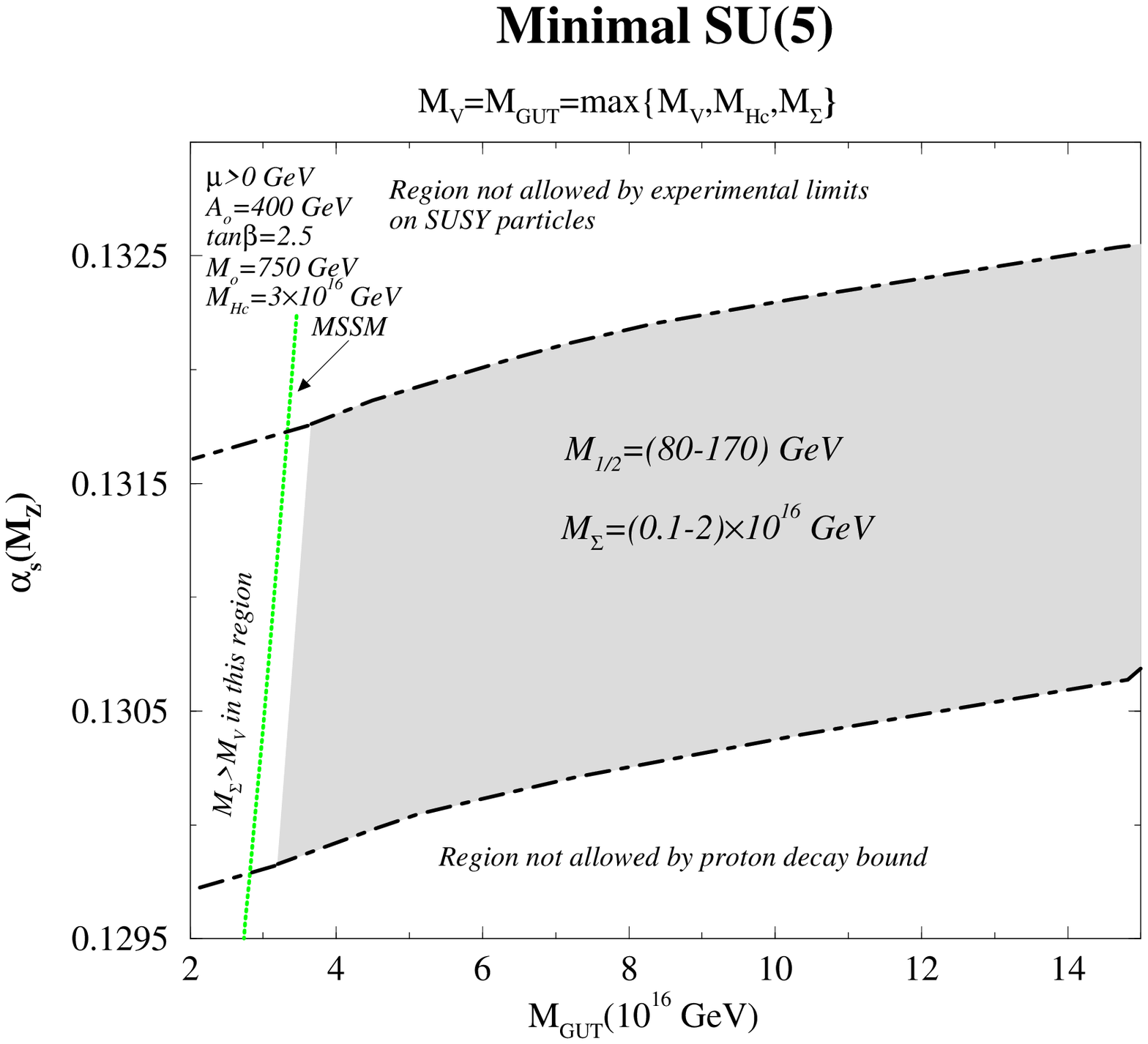}\hss
  }
\caption{\footnotesize (a),(b) $\alpha_s(M_Z)$ versus $M_G$ when $M_{H_c}$,
$M_{\Sigma}$, respectively are varied.}
\end{figure}
Demanding perturbativity up to $\frac{M_{P}}{\sqrt{8\pi}}\simeq
2.4\times 10^{18}
GeV$ for the couplings appearing in (17),
 we are led after numerically integrating the coupled system of $SU(5)$
renormalization group equations, to the inequalities
\begin{equation}
\lambda_1< 1.4\,\,\, ,\lambda_2<1.5 \,\,\, ,  Y_t < 1.5 \,\,\, , Y_b <
1.4
\end{equation}
at $M_G$. Note that although in general
\begin{equation}
M_G=max \{ M_V,M_{H_c},M_{\Sigma} \}\,\,\,\, ,
\end{equation}
only the case $M_G=M_V$ can be realized under the combined constraints
in our analyses. The case in which all superheavy masses are equal although
allowed by the bounds given in Eq. (22) is the well studied case of the MSSM.
In the context of the $SU(5)$ this requires the couplings
 $\lambda_{1,2} ,\ g_5 \,$ to be fine tuned according to (18).

Our standard outputs are the values of the strong coupling,
$\alpha_s(M_Z)$, and
the scale $M_G$ where the couplings meet. The input values of $M_o$
, $M_{1/2}$ and $A_o$ are always taken to be smaller than $1$ TeV.
It must be noted that in the figures
we have chosen input values such that the acceptable region in the
($\alpha_s$-$M_G$) plane is the optimum one in the following sence:
We vary one parameter at a time while we keep the others constant.
This is done for every input parameter, until we reach the maximum
acceptable area.
In the figures shown, we have adopted for the mass of the top quark
an average value of the CDF and D$\O$ $\cite{abe}$
experimental results,
$m_t=180$ GeV.
 Note also that, variation of $\pm 5$GeV in $m_t=180$GeV
results in $\pm 0.0005$ and $\pm 0.08\times 10^{16}$GeV on $\alpha_s$ and
$M_G$ respectively. In addition, a variation of $\pm 0.0007$ in the
central value of $\Delta_{\gamma}$ gives variation of $\mp 0.001$ and
$\mp 0.5\times 10^{16}$GeV on $\alpha_s$ and
$M_G$ respectively.

The shaded areas of Figures 1a and 1b represent the allowed parameter space for
the outputs $\alpha_s(M_Z)$ and $M_G$. The results do not depend significantly
on $M_o$ and $A_o$ which have been chosen to have the representative values
shown. The allowed area shrinks with increasing
$tan\beta$ due to the proton decay bound.
For smaller values the allowed area shrinks because of the
perturbativity of $Y_t(M_G)$ and the constraint from
radiative symmetry breaking. In Figure
1a we have chosen a characteristic value for $M_{\Sigma}$ while we have varied
$M_{H_c}$ through its allowed range of values $(1.8-4.6)\times 10^{16}$ GeV.
Analogously in Figure 1b we have taken $M_{H_c}=3\times 10^{16}$ GeV, while we
have varied $M_{\Sigma}$ through the range of values $(0.1-2)\times
10^{16}$ GeV. The values of $\alpha_s$ obtained are in general too large
in comparison with the average  experimental value. Making a parameter
search, we conclude that
the lowest possible value that
we can reach in this model is close to  $\simeq 0.130$.
 Nevertheless, there are
processes
(Table I) whose determined $\alpha_s$ agrees in a limiting sense with the
smallest values in Fig. 1. It should be noted that the effect of high energy
thresholds has made the access to the smaller values of $\alpha_s$ worse than
in the case of the MSSM. The general dependence of
$\alpha_s$ is that it increases with increasing $M_{H_c}$ while it
decreases with increasing $M_{\Sigma}$.
\vspace{.5 cm}

{\bf  3. Missing Doublet Model}
\vspace{.5 cm}

Let us now consider an extended version of the $SU(5)$ known as the Missing
Doublet Model.
This model \cite{yamada}, constructed in order to avoid the numerical
fine-tuning required in the minimal $SU(5)$, has instead of the adjoint GUT
Higgs a Higgs field in the {\bf{75}} representation as well as an extra pair of
Higgses in the ${\bf {50}}+{\bf{\overline{50}}}$ representation. The
superpotential is
\begin{eqnarray}
\cal W &=& M_{1} Tr(\Sigma^2)+\frac{1}{3} \lambda_{1} Tr(\Sigma^3)+\lambda_2 H
\Sigma\Theta+\overline{\lambda}_2\overline{H}\Sigma\overline{\Theta}+M_{2}
\overline{\Theta}\Theta \nonumber \\[1.5mm] &+& \frac{1}{4} Y_{(u)}^{ij}
\Psi_{i}\Psi_{j}H+\sqrt{2} Y_{(d)}^{ij}\Psi_{i}\phi_{j}\overline{H} \;\;\;\;\;
\; i,j=1..3
\label{potmdm}
\end{eqnarray}
Since the $\Theta$,$\overline{\Theta}$ do not contain any isodoublets, only the
coloured triplets obtain masses while the isodoublets in $H$,$\overline{H}$
stay massless. The v.e.v
\begin{equation}
\Sigma^{CD}_{AB}=V ( (\delta_c)^C_A (\delta_c)^D_B+ 2 (\delta_w)^C_A
(\delta_w)^D_B-\frac{1}{2} \delta^C_A \delta^D_B - (C\leftrightarrow D))
\end{equation}
leads to the masses
\begin{eqnarray}
M(3,1,\frac{5}{3})&=&\frac{4}{5} M_{\Sigma}\,\,\, ,
M(8,1,0)=\frac{M_{\Sigma}}{5} \,\,\, , M(8,3,0)=M_{\Sigma} \nonumber \\[1.5mm]
M(6,2,\frac{5}{6})&=&\frac{2}{5} M_{\Sigma}\,\,\, , M(1,1,0)=\frac{2}{5}
M_{\Sigma} \nonumber
\end{eqnarray}
for the remnants of {\bf 75}. The assignment of the quantum numbers refers
to the group $SU(3) \times SU(2) \times U(1) $ .
We shall assume that the parameter $M_2$ is
larger than the GUT scale possibly of the order of the Planck mass.
Otherwise perturbativity, as can be easily seen, cannot be fulfilled.
The charge -
$1\over 3$ colour triplets in $H$,$\overline{H}$ and
$\Theta$,$\overline{\Theta}$ will give one supermassive combination of mass
$M_{H_{c^{\prime}}}\simeq O(M_2)$ and a light combination of mass
\begin{equation}
M_{H_c}\simeq \frac{9}{100} (\frac{32 \lambda_2
\overline{\lambda}_2}{\lambda_1^2})(\frac{M_{\Sigma}^2}{M_2})
\end{equation}
The mass parameter $M_{\Sigma}$ is related to the vector boson mass through
\begin{equation}
\frac{M_{\Sigma}}{M_V}=\frac{5 M_1}{2 \sqrt{15} g_5 V}=\frac{(10/3)
\lambda_1}{2 \sqrt{15} g_5}=\frac{1}{3} \sqrt{\frac{5}{3}}
(\frac{\lambda_1}{g_5})
\end{equation}
\begin{figure}
\hbox to\hsize{\hss \epsfysize=3in
  \epsfbox[18 18 552 462]{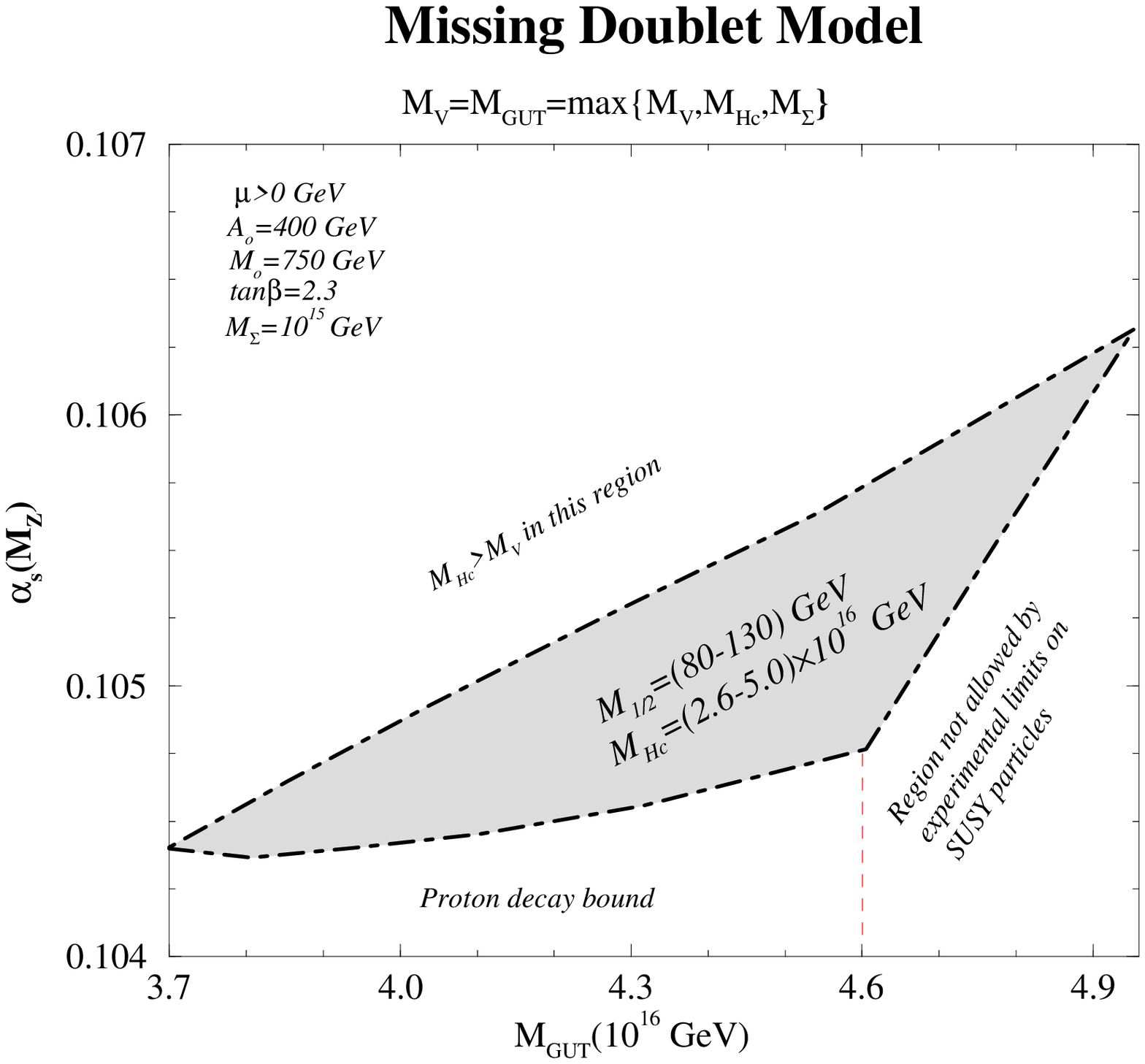}\hss
   \epsfysize=3in\epsfbox[18 18 552 462]{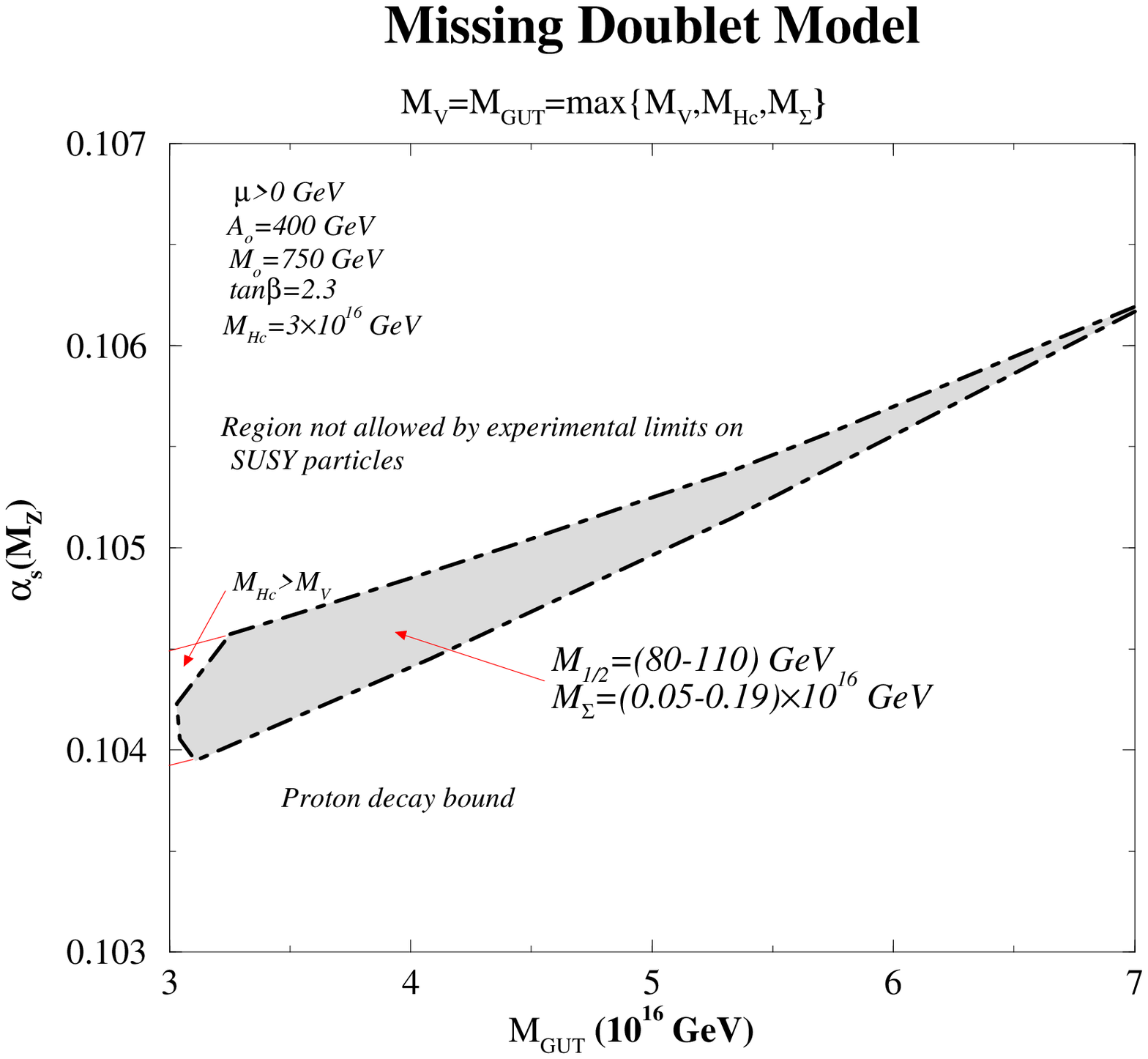}\hss
  }
\caption{\footnotesize (a),(b) $\alpha_s(M_Z)$ versus $M_G$ when $M_{H_c}$,
$M_{\Sigma}$, respectively are varied.}
\end{figure}

The modifications in the beta function coefficients are,
\newpage
\begin{eqnarray}
\Delta b_3 &=& -4\theta(Q^2-M_{V}^2)
+9\theta(Q^2-M_{\Sigma}^2)+
\theta(Q^2-0.8^2M_{\Sigma}^2) \nonumber \\[1.5mm]
&+&10\theta(Q^2-0.4^2M_{\Sigma}^2)+
3\theta(Q^2-0.2^2M_{\Sigma}^2)+\theta(Q^2-M_{H_c}^2)+
\theta(Q^2-M_{H_{c^{\prime}}}^2) \nonumber \\[1.5mm]
&+&34\theta(Q^2-M_{2}^2)\\[2mm]
\Delta b_2 &=& -6\theta(Q^2-M_{V}^2)
+16\theta(Q^2-M_{\Sigma}^2)+6\theta(Q^2-0.4^2M_{\Sigma}^2) \nonumber
\\[1.5mm] &+&35\theta(Q^2-M_{2}^2)\\[2mm]
\Delta b_1 &=& -10\theta(Q^2-M_{V}^2)
+10\theta(Q^2-0.8^2M_{\Sigma}^2)+10\theta(Q^2-0.4^2M_{\Sigma}^2) \nonumber
\\[1.5mm]
&+&\frac{2}{5}\theta(Q^2-M_{H_c}^2)+\frac{2}{5}\theta(Q^2-
M_{H_{c^{\prime}}}^2)+\frac{173}{5}\theta(Q^2-M_{2}^2)
\end{eqnarray}
Perturbativity above $M_G$, as in the case of minimal $SU(5)$, leads us to the
extra constraints at $M_G$

\begin{equation}
\lambda_1< 0.18\,\,\,\, ,  Y_t < 1.6 \,\,\,\, , Y_b < 1.5
\end{equation}
In this model, we obtain a stronger constraint on $\lambda_1$, and consequently
on $M_{\Sigma}$, due to the fact that now we have a larger dimensional
representation ({\bf 75}).

Note that the model as it stands does not contain any $\mu$-term. We assume,
however, that a $\mu$-term is generated through an independent mechanism
$\cite{kim}$.

In Figures 2a and 2b it can be seen that the values of $\alpha_s$ obtained are
much smaller than the average experimental value.
We should note that  $\alpha_s$
is pushed towards smaller values due to the splittings within
{\bf 75} that give a large correction in the opposite direction
than in the case of the minimal model $\cite{yamada}$.
The maximum value of $\alpha_s$ that we are able to obtain in the
{\it Missing Doublet} Model, is approximately $\simeq 0.106$.
As we can see from
Table I, there are still QCD processes, where the values of $\alpha_s$ are
in agreement with the results of the missing doublet model.
The heavy masses $M_{H_c}$ and
$M_{\Sigma}$ are constrained to be in the regions $(2.6-5.0)\times 10^{16}$ GeV
and $(0.05-0.19)\times 10^{16}$ GeV respectively. Due to the fact that the
extracted values of $\alpha_s$ in MSSM are greater than 0.125 (for the inputs
of Figs 2a,b), we do not display in Figures 2a and 2b the corresponding MSSM
plane. Experimental limits on
LSP (see Table II)
puts a lower bound on the universal soft gaugino masses such that
$M_{1/2}\geq 80$ GeV.
\vspace{.5 cm}

{\bf 4. Peccei-Quinn symmetric missing-doublet model}
\vspace{.5 cm}

The problem of proton decay through D=5 operators that is present in the
minimal $SU(5)$ model provided strong motivation to construct versions of
$SU(5)$
with a {\it Peccei-Quinn} symmetry$\cite{quinn}$ that
naturally suppresses these operators by
a factor proportional to the ratio of the {\it Peccei-Quinn} breaking scale
over the GUT scale \cite{tobe}. A {\it Peccei-Quinn} version of the missing
doublet $SU(5)$ model requires the doubling of ${\bf 5}+ {\bf {\overline{5}}}$
and ${\bf 50}+ {\bf {\overline{50}}}$ representations. The relevant
superpotential terms which must be added to the previous
superpotential are,

\begin{figure}
\hbox to\hsize{\hss \epsfysize=3in
  \epsfbox[18 18 552 462]{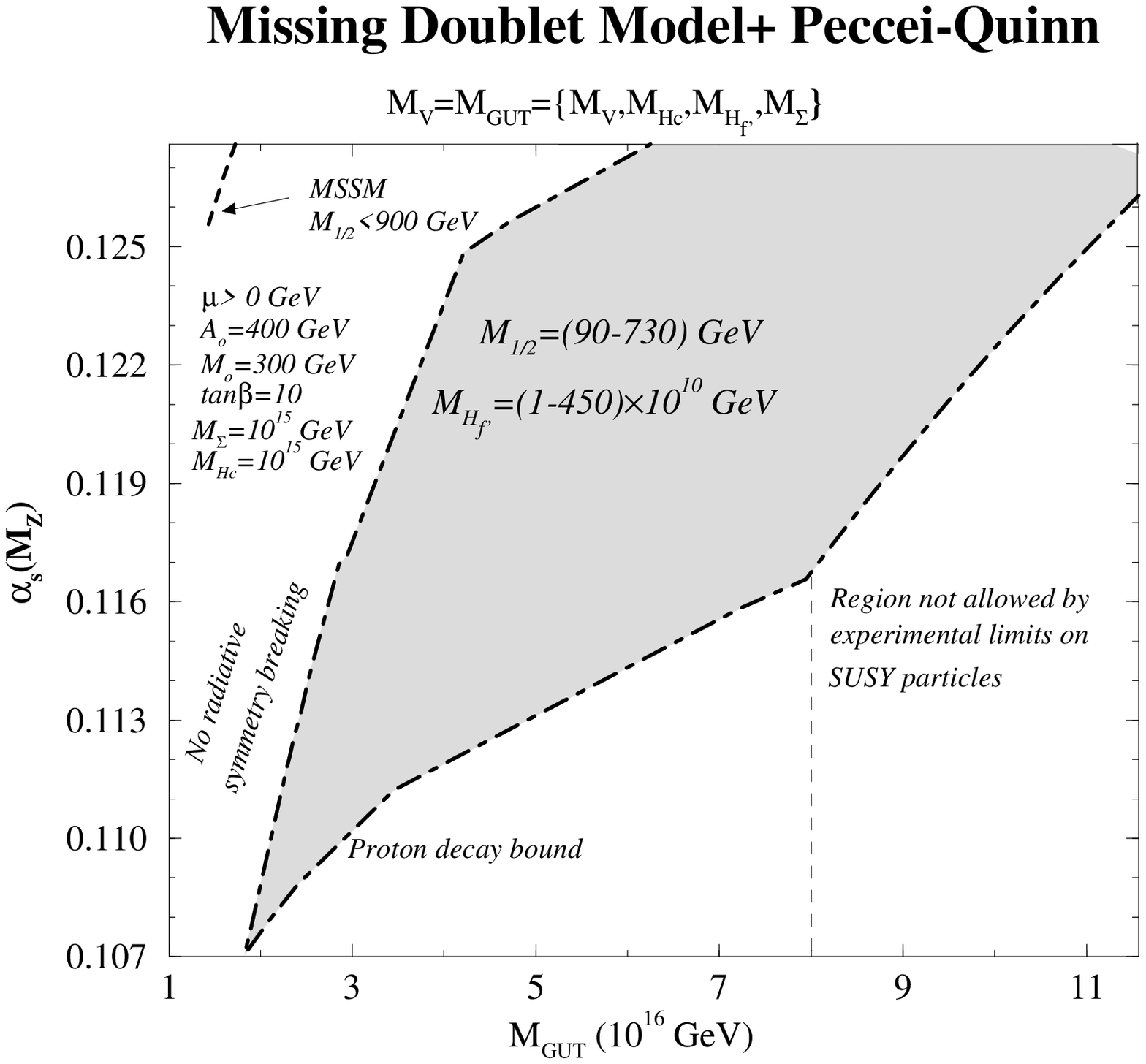}\hss
   \epsfysize=3in\epsfbox[18 18 552 462]{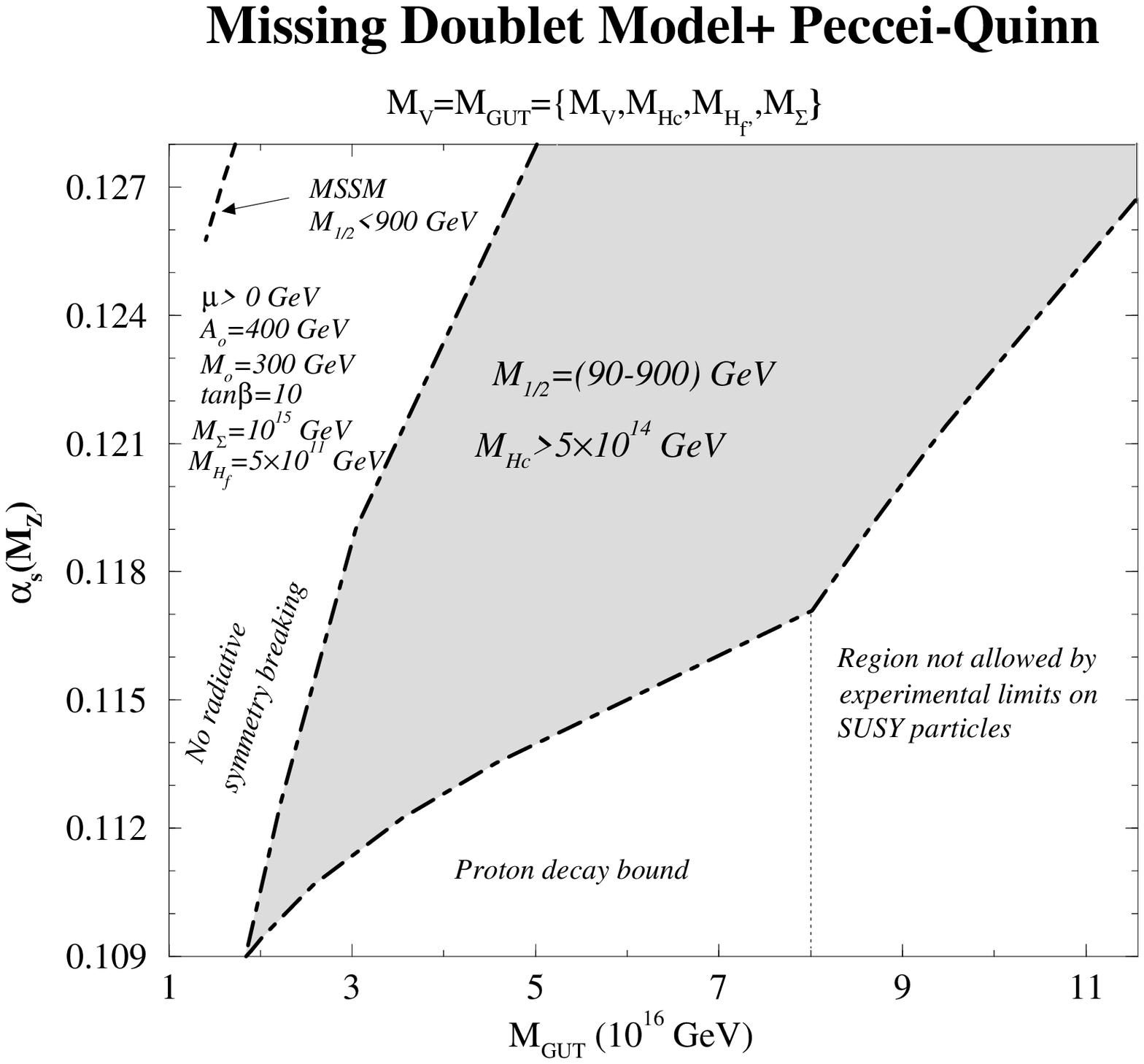}\hss
  }
\hbox to\hsize{\hss \epsfysize=3in
  \epsfbox[18 18 552 462]{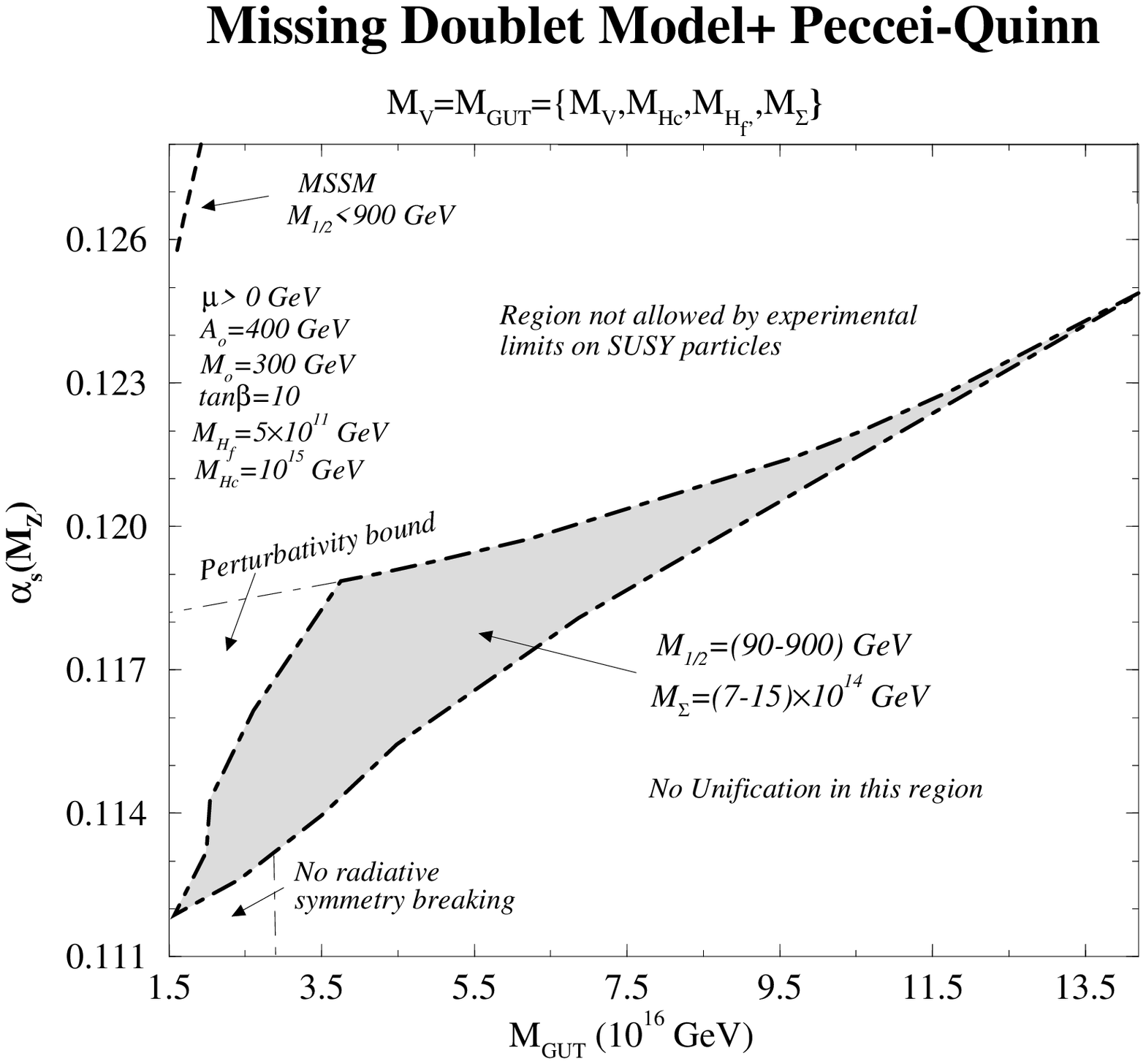}\hss}
\caption{\footnotesize (a),(b),(c) $\alpha_s(M_Z)$ versus $M_G$ when
$M_{H_f^{\prime}}$, $M_{H_c}$ and $M_{\Sigma}$ are varied.}
\end{figure}

\begin{eqnarray}
\lambda_2 H \Sigma \overline{\Theta} &+& \overline{\lambda}_2 \overline{H}
\Sigma \Theta + \lambda^{\prime}_2 H^{\prime} \Sigma
\overline{\Theta}^{\prime}+\overline{\lambda}^{\prime}_2 \overline{H}^{\prime}
\Sigma \Theta^{\prime} \nonumber \\[2mm]
&+& M_2 \Theta \overline{\Theta}^{\prime}+ M_2^{\prime} \Theta^{\prime}
\overline {\Theta} + \lambda_3 P \overline{H}^{\prime} H^{\prime}
\end{eqnarray}
$P$ stands for an extra gauge singlet superfield. The charges under the {\it
Peccei-Quinn} symmetry are $\Psi (\alpha/2)$, $\phi (\beta/2)$, $H(-\alpha)$,
$\overline{H}^{\prime}(-\frac {\alpha + \beta}{2})$,
$\overline{\Theta}(\alpha)$, $\Theta(\frac {\alpha + \beta}{2})$,
$\Theta^{\prime}(-\alpha)$, $\overline{\Theta}^{\prime}(-\frac {\alpha +
\beta}{2})$, $H^{\prime}(\frac {\alpha + \beta}{2})$,
$\overline{H}^{\prime}(\alpha)$, $P(-(3\alpha +\beta))$. The breaking of the
{\it Peccei-Quinn} symmetry can be achieved with a suitable gauge singlet
system at an intermediate energy $<P> \equiv \frac{M_{H^{\prime}_f}}{\lambda_3}
\sim 10^{10}-10^{12} GeV$. Assuming $M_2$, $M_2^{\prime}$ to be of the order of
the Planck scale, we obtain two massive pairs of coloured triplets with masses
\begin{equation}
M_{H_c} \simeq 32 \lambda_2 \overline{\lambda}^{\prime}_2 \frac {V^2}{M_2}
 \;\;\;,\;\;\; M_{\overline{H}_c} \simeq 32 \overline{\lambda}_2
\lambda^{\prime}_2
\frac{V^2}{M_2^{\prime}}
\end{equation}
somewhat below the GUT scale. Note that $M_V=2 \sqrt{15} g_5 V$. In addition we
have two pairs of isodoublets one of which is massless while the other pair
receives the intermediate mass $M_{H^{\prime}_f}$. The modifications in the
renormalization group beta functions coefficients are,
\begin{eqnarray}
\Delta b_3 &=& -4\theta(Q^2-M_{V}^2)
+9\theta(Q^2-M_{\Sigma}^2)+
\theta(Q^2-0.8^2M_{\Sigma}^2) \nonumber \\[1.5mm]
&+&10\theta(Q^2-0.4^2M_{\Sigma}^2)+
3\theta(Q^2-0.2^2M_{\Sigma}^2)+\theta(Q^2-M_{H_c}^2)+
\theta(Q^2-M_{\overline{H}_{c}}^2) \\[2mm]
\Delta b_2 &=&-6\theta(Q^2-M_{V}^2)
+16\theta(Q^2-M_{\Sigma}^2)+6\theta(Q^2-0.4^2M_{\Sigma}^2) \nonumber
\\[1.5mm] &+&\theta(Q^2-M_{H_f^{\prime}}^2)\\[2mm]
\end{eqnarray}
\begin{eqnarray}
\Delta b_1 &=& -10\theta(Q^2-M_{V}^2)
+10\theta(Q^2-0.8^2M_{\Sigma}^2)+10\theta(Q^2-0.4^2M_{\Sigma}^2) \nonumber
\\[1.5mm]
&+&\frac{2}{5}\theta(Q^2-M_{H_c}^2)+
\frac{2}{5}\theta(Q^2-M_{\overline{H}_c}^2)+
\frac{3}{5}\theta(Q^2-M_{{H}_f^{\prime}}^2)
\end{eqnarray}

Bounds, which come from perturbativity of couplings in this model, are
numerically similar, to those of the previous one
(see Appendix). The extra coupling
$\lambda_3$ obeys the constraint $\lambda_3 < 2.7$ at
$M_G$.

It is evident from figures 3 that the values of $\alpha_s$ obtained for the
case of this model are in excellent agreement with the experiment.
The range of these values $(0.107-0.140)$, covers the experimental average
value of
$\alpha_s$ and lays between the gap
of the {\it Minimal SU(5)} model and the {\it
Missing Doublet} model.
This model
possesses an additional parameter, the intermediate scale $M_{H_f^{\prime}}$
which increases the values of $\alpha_s$ when it takes lower values. The
allowed range of values for $M_{1/2}$ has been increased now to 730 GeV or 900
GeV in Figure 3a and 3b respectively. For this model the allowed range for
$M_o$ can be extended to lower values. Figure 3a and 3b have been obtained for
$M_o=300$ GeV. Note however that still $\alpha_s$ does not depend significantly
on $M_o$. In addition, $tan\beta$ can practically now take much larger values,
as large as $tan\beta\sim 40$. Figures 3a, 3b and 3c have been obtained for an
intermediate value $tan\beta=10$. The allowed range of values for the
intermediate scale $M_{H_f^{\prime}}$ is $(1-450)\times 10^{10}$ GeV .
Similarly $5\times 10^{14}<M_{H_c}<M_G$ and $7 \times 10^{14}<M_{\Sigma}<15
\times 10^{14}$ GeV . Note finally, that the grand unification scale can take
values as large as the ``string scale" for rather large but not excluded
values of $\alpha_s$.
\vspace{.5 cm}

{\bf 5. A version of SU(5) with light remnants }
\vspace{.5 cm}

Recently there has been some activity around models with gauge group $G\times
G$ with intent to
 bypass the known problem of k=1 superstring constructions where no
adjoint Higgs can appear in the massless spectrum \cite{bachas}. Vector-vector
Higgses present in the spectrum of $G\times G$ can break it into $G_{diag}$. An
$SU(5)\times SU(5)$ model with Higgses
$Z(5,\overline{5})+\overline{Z}(\overline{5},5)$ can have renormalizable
couplings only of the type $\Phi Z^i_j \overline{Z}^j_i=\Phi Tr(Z\overline{Z})$
to a singlet $\Phi$, in addition to self-couplings of singlets. Thus we could
construct an analog $SU(5)$ GUT with superpotential,
\begin{figure}
\hbox to\hsize{\hss \epsfysize=3in
  \epsfbox[18 18 552 462]{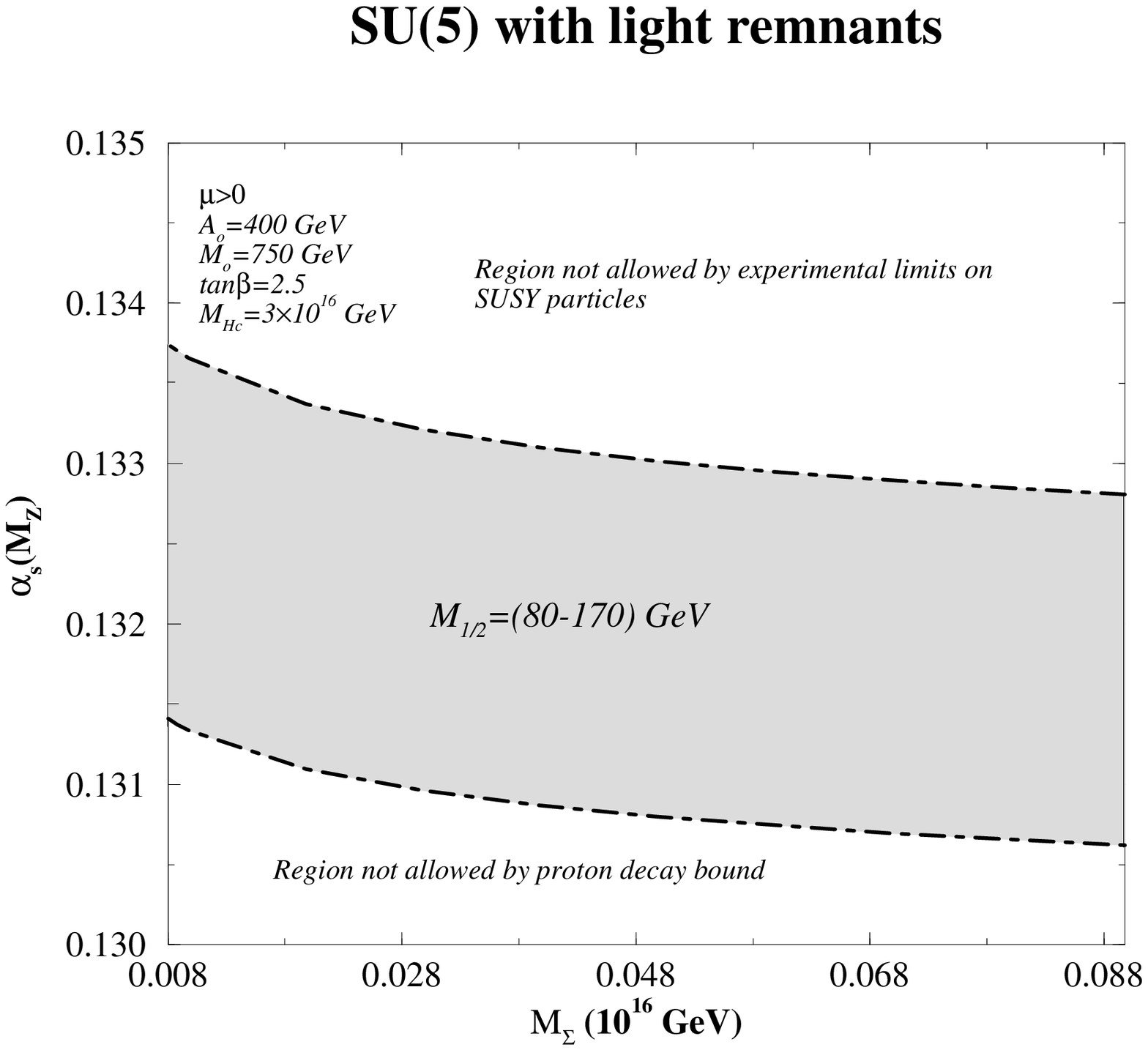}\hss
   \epsfysize=3in\epsfbox[18 18 552 462]{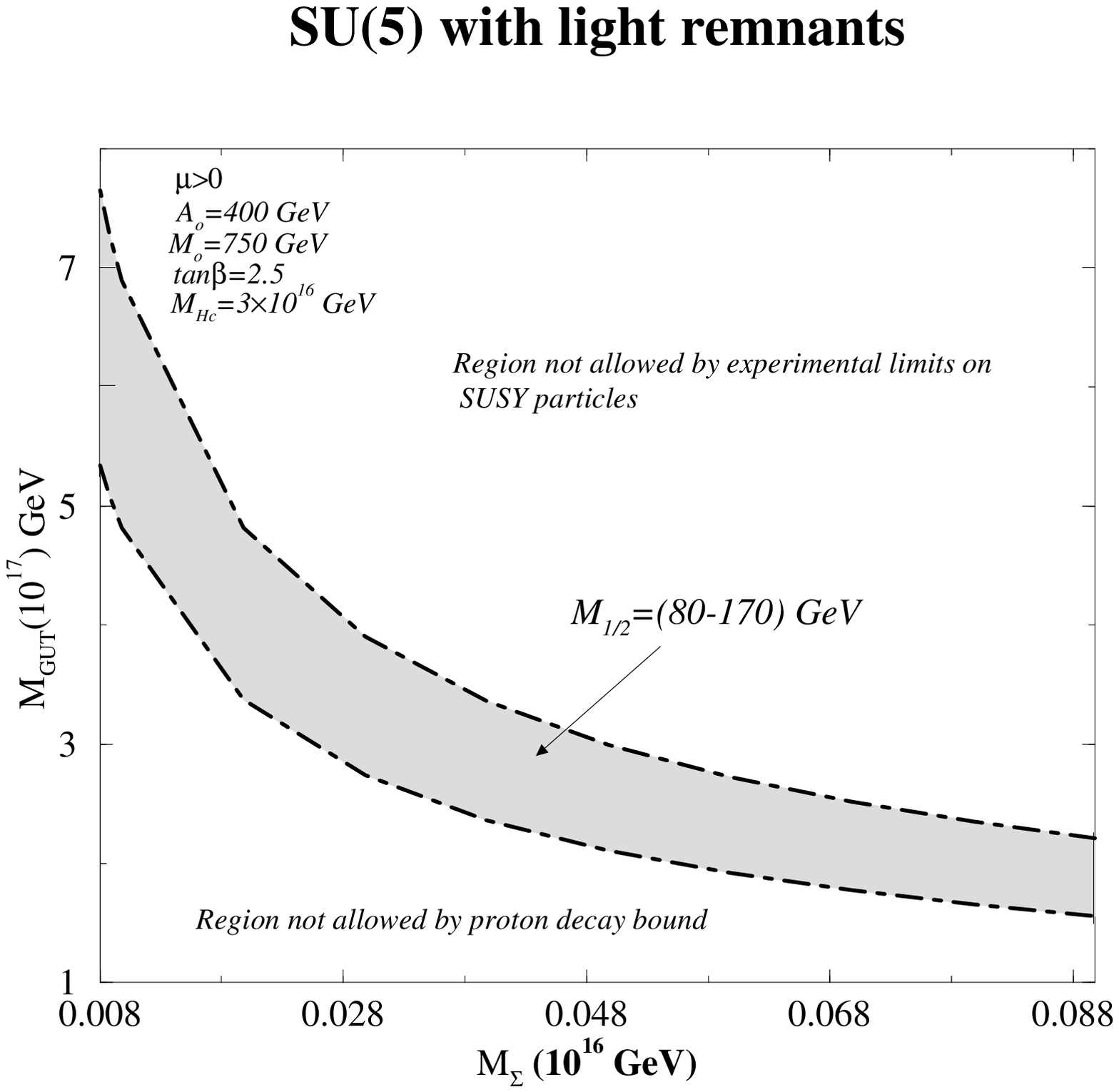}\hss
  }
\caption{\footnotesize (a),(b) $\alpha_s$ and $M_G$
vs $M_{\Sigma}$ in SU(5) model with light remnants.}
\end{figure}
\begin{equation}
{\cal W} = \frac{\lambda_1}{2} \Phi_1 Tr(\Sigma^2)+\frac{\lambda_2}{2} \Phi_1
\Phi_2^2 +\frac{\lambda_3}{3} \Phi_1^3
\end{equation}
where $\Sigma$ is the adjoint and $\Phi_1$, $\Phi_2$ are singlets. This
superpotential is invariant under $\Phi_2\rightarrow -\Phi_2$. The F-flatness
conditions give, apart from $<\Sigma>=V Diag(2,2,2,-3,-3)$,
\begin{equation}
\Phi_2^2=-\frac{\lambda_1}{\lambda_2}(30 V^2) \,\,\,\, , \Phi_1=0
\end{equation}
With the above superpotential the remnants of $\Sigma$, a colour octet and an
isotriplet, stay massless. Nevertheless, non-renormalizable terms like
\begin{equation}
\Delta {\cal W}=\frac{\lambda_4}{M} Tr(\Sigma^4)+\frac{\lambda_5}{M}
(Tr(\Sigma^2))^2
\end{equation}
can in principle induce a mass of order $O(\frac{V^2}{M})\sim 10^{14} GeV$ or
smaller, depending on the actual values of $\lambda_4$ and $\lambda_5$.
A higher order
non-renormalizable term would induce an even smaller mass
$\frac{V^3}{M^2}\sim
10^{12} GeV$. The light $\Sigma$ in this model could allow for a large
$M_G$ close to a string unification scale
and thus in such a model there would
be no string unification mismatch.

We shall therefore investigate
the effects of small $M_{\Sigma}$ on $\alpha_s$ and $M_G$.
In order to obtain acceptably small
values of $\alpha_s$, we choose as input value of $M_{H_c}$
the smallest acceptable one as it is shown in Figures 4a and
4b, since an increasing  $M_{H_c}$ tends to increase $\alpha_s$.
When we vary
$M_{\Sigma}$ from $10^{15}$ down to $10^{13}$ GeV,
we can achieve unification at
$M_G=5\times 10^{17}$ GeV $\simeq O(M_{string})$.
Decreasing $M_{\Sigma}$ further towards the intermediate scale,
leads to even larger values of $\alpha_s$.
However, the values of $\alpha_s$ are still
rather large ($>0.131$). This excludes this particular
version at least in this simple form.
\vspace{.5 cm}

{\bf 6. Conclusions}
\vspace{.5 cm}

In this article we have studied various supersymmetric
extensions of the Standard Model based on the group $SU(5)$.
The low energy
precision data in conjunction with the existing experimental bounds
on sparticle masses are known to impose strong constraints if
radiative breaking of the electroweak symmetry is assumed.
There exist several detailed studies in the
literature in the framework of
radiative symmetry breaking
$\cite{Ross,castano,vayonakis,bagger,dedes,zhang}$,
which however have not considered in detail
the effect of the superheavy degrees of freedom
included in unified schemes. In SUSY GUTs there
are additional constraints one has to deal with such as the experimental
bound on proton's lifetime, the absence of Landau poles beyond the
unification scale and the appearance of heavy thresholds which influence
the evolution of the couplings involved. All these affect the low
energy predictions. The existing analyses in this direction on the other
hand $\cite{nath,hisano,yamada,wright,tobe,bachas}$,
have not systematically taken into account the effect of the low
energy thresholds at the level of accuracy required by
low energy precision
data as was done in the previous references.
Our analysis combines both
and takes into account high and low energy thresholds at the accuracy
required by precision experiments. In particular we have focused our
attention on the extracted value of the strong coupling constant
$\alpha_s(M_Z)$ , the value of the unification scale $M_G$, as well as
the restrictions imposed on the heavy masses in some unifying schemes
based on the $SU(5)$. Sample results of our findings have been displayed
in figures 1-4.

In the case of the minimal $SU(5)$  we found that the values of the
strong coupling constant obtained are somewhat larger as compared to
the average experimental value of $\alpha_s(M_Z)$. Also the unification
scale $M_G$ differs from the string unification scale
by an order of magnitude if the lower values of $\alpha_s(M_Z)$ obtained
are assumed. Access to small values of the strong coupling constant
is more difficult than in the MSSM exhibiting the influence of the
superheavy degrees of freedom in a clear manner. The range  of
$M_{H_{c}}$ allowed in this model is somewhat limited.
The {\it Missing doublet model} seems to favour small values of
$\alpha_s(M_Z)$, in contrast to those obtained in the MSSM and the
minimal $SU(5)$ version. At the same time when $M_G$ increases
the allowed parameter space shrinks considerably.
Proton decay along with
perturbativity requirements seem to put a stringent constraint on both
minimal SU(5) and Missing Doublet Model (MDM). In the MDM the large
splittings within $\bf {75}$ give a high energy threshold effect on
$\alpha_s(M_Z)$ in the opposite direction than in the case of the minimal
model leading to small values. This is also the case for the
{\it Peccei-Quinn} version of the MDM. However the presence of an extra
intermediate scale ameliorates the situation allowing to achieve an
excellent agreement with the experimental values of $\alpha_s(M_Z)$.
The grand unification scale can take values as large as the string scale
at the expense however of having rather large values of the strong coupling
constant not favoured by all experiments.
If we consider the range for allowed values of $\alpha_s(M_Z)$,
 the minimal
SU(5) lays always above $.130$ while MDM lays below $.106$. The
intermediate range between the two can be covered by the
{\it Peccei-Quinn} version of the MDM
and coincides with the allowed experimental range.
Finally, the last considered version, with light $\Sigma$, exhibits
unification of couplings at string scale but the values of $\alpha_s$
obtained are rather large, although within the errors of some
experiments.
\vspace{.5 cm}

{\bf Acknowledgements}
\vspace{.5 cm}

A.D. would like to thank J. Rosiek for useful conversations.
A.B.L. acknowledges support by the EEC Science
Program No. SCI-CT92-0792. A.D. acknowledges support from the
Program $\Pi$ENE$\Delta$ 95, the EEC Human Capital and Mobility
Program CHRX-CT93-0319 and the CHRX-CT93-0132 (``Flavordynamics").

\newpage
{\bf Appendix :  RGE's above $M_G$}

The Renormalization group equations for the gauge and Yukawa couplings from
$M_{GUT}$ to $\frac{M_{P}}{\sqrt{8\pi}}\simeq 2.4\times 10^{18}
GeV$ in the case of minimal $SU(5)$ are \cite{hisano,wright},
\begin{eqnarray}
\frac{dg_{5}}{dt}&=&-\frac{3}{(4\pi)^2}g_{5}^3 \\[2mm]
\frac{d\lambda_{1}}{dt}&=&\frac{\lambda_{1}}{(4\pi)^2}
(\frac{63}{5}\lambda_{1}^2+3\lambda_{2}^2-30g_{5}^2) \\[2mm]
\frac{d\lambda_{2}}{dt}&=&\frac{\lambda_{2}}{(4\pi)^2}
(\frac{21}{5}\lambda_{1}^2+\frac{53}{5}\lambda_{2}^2
-\frac{98}{5}g_{5}^2+3 Y_t^2+4 Y_b^2)\\[2mm]
\frac{dY_t}{dt}&=&\frac{Y_t}{(4\pi)^2} (\frac{24}{5}\lambda_{2}^2+9 Y_t^2+
4 Y_b^2-\frac{96}{5} g_{5}^2)\\[2mm]
\frac{dY_b}{dt}&=&\frac{Y_b}{(4\pi)^2} (\frac{24}{5}\lambda_{2}^2+3 Y_t^2+
10 Y_b^2-\frac{84}{5}g_{5}^2)
\end{eqnarray}
where  $t=ln(\frac {Q}{M_P/\sqrt{8\pi}})$.

The modification of the above system in the case of the missing doublet model
is,
\begin{eqnarray}
\frac{dg_{5}}{dt}&=&\frac{17}{(4\pi)^2}g_{5}^3 \\[2mm]
\frac{d\lambda_{1}}{dt}&=&\frac{\lambda_{1}}{(4\pi)^2}
(\frac{56}{3}(8\lambda_{1})^2-48g_{5}^2)
\end{eqnarray}
The other two equations for the top and bottom Yukawa coupling are derived if
we set $\lambda_2=0$ in the eqs.(43,44) of the minimal model. Due to the fact
that we have an extra pair of Higgs ${\bf{5}}$ and ${\bf{\overline{5}}}$ in the
Peccei-Quinn version of the missing doublet model, the only equation that
changes compared to the missing doublet model is the one for the gauge coupling
which takes the form
\begin{equation}
\frac{dg_{5}}{dt}=\frac{18}{(4\pi)^2}g_{5}^3
\end{equation}
In this model we have an extra coupling $\lambda_3$ whose running is
given by the following renormalization group equation
\begin{equation}
\frac{d\lambda_{3}}{dt}= \frac{\lambda_{3}}{(4 \pi)^2}
(3 \lambda_{3}^2 - \frac{48}{5} g_{5}^2)
\end{equation}

\newpage

\end{document}